\documentclass[twocolumn,twoside,slac_two]{revtex4}
\usepackage{graphicx}
\usepackage{fancyhdr}

\def\be{\begin{equation}}
\def\ee{\end{equation}}
\def\bea{\begin{eqnarray}}
\def\eea{\end{eqnarray}}

\begin{document}

\title{Neutrino oscillations in a variable-density medium and 
$\nu-$ bursts due to the gravitational collapse of stars}

\author{S. P. Mikheev and A. Yu. Smirnov}

\null
\vspace*{0.3cm}

\affiliation{Institute of Nuclear Research, Academy of Sciences of the USSR\\
(Submitted 24 December 1985)\\
Zh. Eksp. Teor. Fiz. {\bf 91}, 7-13 (July 1986)}

\vspace*{0.1cm}

\begin{abstract}

Under certain conditions, the propagation of a beam of oscillating 
neutrinos in a variable-density medium takes the form 
of an almost complete transformation of the initial type of 
neutrino into another type. The depth of oscillations is then negligible. 
The transformation can occur in the cores and envelopes of 
collapsing stars. 

\end{abstract}

\maketitle

\thispagestyle{fancy}

\section{Introduction}
 
The interaction of neutrinos and matter modifies the picture 
of $\nu$-oscillations \cite{W1,W2,bar}. The effect of the medium is 
analogous in a number of significant respects to the coherent
regeneration of $K$ mesons, and also to the appearance of a
refractive index.

In a medium of constant density, the overall character of the 
oscillations is the same as in vacuum, i.e., there is a
change in only the oscillation length and depth. A number of
astrophysical \cite{W1,pak,haub} and geophysical \cite{W1,bar,ram} 
applications of these oscillations have been considered in the constant-density 
approximation. 

In the present paper, we examine qualitatively new 
effects that appear in a variable-density medium.

\section{Equation for the transformation probability;
resonance condition}

We shall consider the mixing of two types of neutrino.
To be specific, we shall suppose that they are $\nu_e$ and $\nu_{\mu}$.
Suppose the $\nu_e$ are created in the source. We shall seek the
probability $P(t)$ of an oscillatory $\nu_e\rightarrow\nu_e$
transformation in a time $t$ (or distance $r\sim ct$ from the source).
The equations describing the evolution of the wave functions of the
$\nu_e$ and $\nu_{\mu}$ in the medium \cite{W1} (essentially, the
Schroedinger equations) can be used to show that the probability
$P(t)$ is the solution of the following equation:
\begin{eqnarray}
M \frac{d^3P}{dt^3} - \frac{dM}{dt}\frac{d^2 P}{dt^2} + 
M (M^2 + 4\bar{M}^2) \frac{d P}{dt} -\\ 
\nonumber
- 2\bar{M}^2 \frac{dM}{dt} (2P-1)=0,                  
\label{eq1}
\end{eqnarray}
where 
\be
M=(2\pi / l_{\nu})(\cos2\theta - l_{\nu}/l_{0}),~~~
2\bar{M} = (2\pi/l_{\nu})\sin 2\theta,                           
\label{eq2}
\ee
$l_{\nu}=4\pi E/\Delta m^2$ is the oscillation length in vacuum, $E$ is
the neutrino energy, $\Delta m^2 = m^2_1 - m^2_2$ 
is the difference between
the squares of the masses, $\theta$ is the mixing angle, $l_0$ is a 
characteristic length of the medium, given by
\be
l^{-1}_0 = \frac{\rho}{m_N} \sum_i X_i \frac{\Delta f_i(0)}{2\pi E} 
\label{eq3}
\ee
$\Delta f_i(0) = f_{ie}(0) - f_{i\mu}(0)$ is the difference between 
the $\nu_e$ and $\nu_{\mu}$ forward-scattering amplitudes for the 
$i$th component of the medium $(i=e,p,n)$, $X_i$ is the abundance of the 
$i$th component per nucleon, $\rho$ is the density of the medium, and
$m_{N}$ is the nucleon mass. In accordance with (3), we have
$l^{-1}_0 \sim G_{F} \rho/ m_N$, 
where $G_F$ is the Fermi constant.

The initial conditions for (1) are:
\be
P(0)=1, ~~\frac{dP(0)}{dt} =0, ~~\frac{d^2 P}{dt^2} = -2\bar{M}^2.                                   
\label{eq4}
\ee
If the density is constant $(dM/dt=0)$, we find from (1) that
\be
\frac{d^3 P}{dt^3}  + (M^2+4\bar{M}^2) \frac{dP}{dt} = 0.
\ee
This equation has a periodic solution of the form
$$
P= 1-A \sin^2(\pi r/l_m)
$$
with oscillation length
\be
l_m = 2\pi \left(M^2 + 4\bar{M}^2 \right)^{-1/2}                                 
\label{eq6}
\ee
and oscillation depth determined by the mixing angle in the
medium~\footnote{Apart from $\theta_m$, the oscillation 
depth is also found to depend on the
initial conditions. The formula (7) 
then corresponds to the initial condition (4).}: 
\be
A = \sin^2 2\theta_m = \bar{M}^2 l_m^2/ \pi^2.           
\label{eq7}
\ee
We recall that the mixing angle $\theta_m$ relates the $\nu_e$ and
$\nu_{\mu}$ states with the neutrino eigenstates $\nu_1^m$ and $\nu_2^m$
in the medium. 
(In the medium the states with the specific masses  $\nu_1$ and $\nu_2$
are not eigenstates of the Hamiltonian and themselves oscillate).

Using (2) and (6), we can rewrite (7) in the form
\be
\sin^2 2\theta_m = \sin^2 2 \theta[(\cos2 \theta - l_{\nu}/l_0)^2 + \sin^2 2 \theta]^{-1} 
\label{eq8}
\ee
from which it follows that, for low values of $\sin^2 2\theta$, the
dependence of $\sin^2 2\theta_m$ on $l_{\nu}/l_0 (l_{\nu}/l_0 \sim \rho E)$
exhibits a resonance. When
\be
l_{\nu}/l_0= \cos 2\theta,
\label{eq9}
\ee
$\sin^2 2\theta_m$ is a maximum, i.e., $\sin^2 2\theta_m = 1$. The quantity
$\sin^2 2 \theta_m$ falls rapidly as $l_{\nu}/l_0$ departs from $\cos2\theta
\equiv 1$. Condition (9), for which the mixing angle in the medium is
equal to $45^{\circ}$ $(\sin^2 2\theta_m = 1)$, will be referred to as the resonance
condition. Correspondingly, the values of $\rho$ and $E$ for which (9)
is satisfied will be referred to as the resonance values. From (9) and
(3), we have
\be
\rho_R= m_N \Delta m^2 \cos2 \theta/ 2^{3/2} G_{F}E.
\label{10}
\ee
The width $\Delta \rho_R$ of the resonance layer will be defined as the
density interval around $\rho_R$ in which $\sin^2 2 \theta_m > {1/2}$.
From (8) we have
\be
\Delta \rho_R=\rho_R \tan 2 \theta.
\label{11}
\ee
The resonance becomes narrower as the mixing angle decreases. Similarly,
we may introduce a resonance energy $E_R$ and a resonance width
$\Delta E_R = E_R \tan2\theta$.

The physical meaning of the resonant behavior of $\sin^2 2 \theta_m$ is as
follows. Suppose that a constant-density layer intercepts a neutrino
flux with a continuous energy spectrum. Neutrinos with energy $E = E_R(\rho)$
will then oscillate with maximum oscillation depth (despite the small
mixing in vaccum). The energy dependence of the oscillation amplitude is
determined by the resonance curve (8): $\sin^2 2 \theta_m = f[l_{\nu}(E)]$.

In the case of a variable-density medium, the oscillation depth does
not depend on $\theta_m$ alone, and even in the resonance layer
$(\rho=\rho_R)$, it may turn out to be low (see below).

\section {Medium with slowly-varying density (adiabatic regime);
nonoscillatory transformation in the $\nu$-beam}

Consider a layer of a medium satisfying the following conditions:

(1) The density distribution $\rho(r)$ has no singularities and
the derivative $d\rho/dr$ is a smooth function.

(2) The variation of density with $r$ is relatively slow and such
that, for a given mixing angle, the width $\Delta r_R$ of the resonance
layer in the $r$ scale is larger than the oscillation length
\be
\Delta r_R = (d\rho/dr)^{-1} \Delta \rho_R  \geq l_m^R/2.
\label{eq12}
\ee
This will be referred to as the adiabatic condition.

(3) The density distribution $\rho(r)$ is a monotonic function of
$r$ up to $\rho_{min} \approx 0$. We shall also assume that the
resonance density for neutrinos of given energy $E$ falls into the
interval $\rho_0 - \rho_{min}$, so that the neutrinos generated in the
region with $\rho = \rho_0$ will traverse the resonance layer.

Under these conditions, the solution $P(r)$ is universal with respect
to the density distribution. The universality can conveniently be
expressed in terms of the following dimensionless parameters. 
Let
\be
n=[\rho(r)-\rho_R] / \Delta \rho_R,
\label{13}
\ee
where $\rho_R$ and $\Delta \rho_R$ are determined by the values of
$E$, $\theta$, and $\Delta m^2$. We shall measure distance in units of $n$ 
rather than $r$. We note that, at resonance, $n=0$, and, as 
$\rho \rightarrow \infty$, $n \rightarrow\infty$, while for 
$\rho \rightarrow0, n\rightarrow - (\tan 2\theta)^{-1}$. 
The initial conditions are set at
\be
n_0 =  [\rho(r_0)-\rho_R]/\Delta \rho_R,
\label{eq14}
\ee
where $n_0$ is the number of resonance layers that can be fitted
between the point at which the neutrino is created and the resonance
layer. When $\rho(r_0) \gg \rho_R$, we have
$$
n_0 \approx \rho(r_0)/\rho_R \sin 2 \theta \approx 1/ \sin2\theta_m^0
$$
[see also (8)], i.e., $n_0$ is equal to the reciprocal of the mixing
parameter at the point at which the $\nu$ is created, and increases
with distance from the resonance layer and/or with decreasing
$\sin^2 2\theta$. 

We also introduce the further variable
\be
m = (d\rho/dr)^{-1}\Delta \rho_R/l^R_m,
\label{eq15}
\ee
i.e., the number of oscillation lengths in the resonance layer.
This number increases as the distribution $\rho(r)$ becomes shallower
or $\sin^2 2 \theta$ increases.

When conditions (1)-(3) are satisfied, the solution $P(r)$ is a 
function of $n$, $n_0$ and $m$, i.e., $P \sim P(n, n_0, m)$, 
and is not very dependent on the density distribution $\rho(r)$. 

Let us examine some of the properties of the function $P(n,n_0,m)$:

(a) $P$ is an oscillating function of $n$ with period $T$; by
definition, $T\sim1/m$ in the resonance layer and, as we depart
from resonance, $T \sim 1 / [m (n^2+1)^{1/2}]$.

(b) $P$ oscillates around its mean value
\be
\bar{P}(n, n_0)=[1 + n_0(n^2_0 + 1)^{-1/2} n (n^2 + 1)^{-1/2}]/2;
\label{eq16}
\ee

(c) The amplitude of the $P$ oscillations is $A_p=|P_{max}-P_{min}|$ and
is a maximum in the resonance layer
\be
A^R_P\approx(n^2_0+1)^{-1/2}
\label{17}
\ee
and decreases with distance from the layer. $A_P$ is practically
independent of $m$.

(d) At exit from the layer, for $\rho\rightarrow0$,
\be
\bar{P}_0 \rightarrow [1-n_0(n^2_0+1)^{-1/2} \cos 2 \theta]/2 .
\label{18}
\ee
Consider the limit of large $n_0$. Increasing $n_0$ means that the
point at which the $\nu$ is created becomes more distant (in $\rho$)
from the resonance layer. The oscillation amplitude is then shown
by (17) to decrease $(A_P \sim 1/n_0)$ and the average of $P(n,n_0)$
tends to the asymptotic value given by
\be
\bar {P}_a(n)=[1 + n(n^2+1)^{-1/2}]/2.
\label{19}
\ee
At exit, $\bar{P}_a \rightarrow \sin^2 \theta$ (see Fig. 1).

\begin{figure}[htb]
\centering
\hspace*{-0.50cm} 
\includegraphics[width=53mm,angle=90]{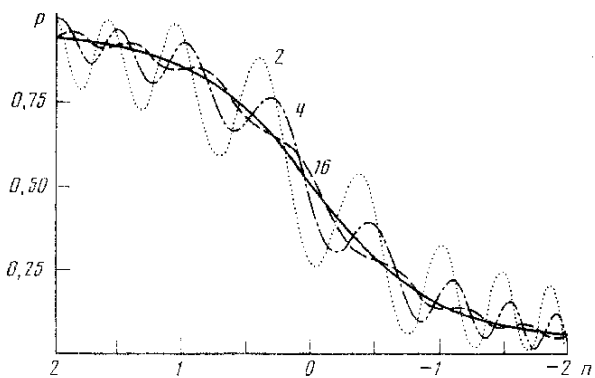}
\caption{~Oscillation probability $P(n)$  for different values of  $n_0$ 
(indicated against the curve).  Solid curve - asymptotic form of 
$\bar{P}(n)$ (nonoscillatory transformation).}
\label{figg1}
\end{figure}

Thus, for small mixing angles $\theta$ in vacuum and large $n_0$
(so that the $\nu$-flux is generated quite far from the resonance
layer), the neutrino propagation process takes the form of a
virtually nonoscillatory transformation of one type of neutrino
into the other.

The interpretation of these results is as follows. The neutrino 
oscillations in the medium take place around the eigenstates
$\nu^m_1$ and $\nu^m_2$, i.e., the neutrino oscillates, as in the
vacuum, around states of specific mass of $\nu_1$ and $\nu_2$.
If the density of the medium varies, there is also a variation in the
eigenstates $\nu^m_1$ and $\nu^m_2$ or, more precisely, there is a 
change in the mixture of $\nu_e$ and $\nu_{\mu}$. When $\rho\gg
\rho_R$, it can be shown from (8) that $\theta_m \sim \pi/2$, i.e.,
$\nu^m_2$ becomes practically identical with $\nu_e$. When $\rho=
\rho_R$, we have $\theta_m=\pi/4$ and $\nu^m_2$ contains equal
admixtures of $\nu_e$ and $\nu_{\mu}$. When $\rho \ll \rho_R$, we
have $\theta_m= \theta$ and, when $\theta$ is small, $\nu^m_2$ consists
mostly of $\nu_{\mu}$. When the density changes from $\rho\gg 
\rho_R$ to $\rho\ll\rho_R$, the basis $\nu^m_1,\nu^m_2$ rotates
through $90^{\circ}$ relative to the basis $\nu_e,\nu_{\mu}$. The adiabatic
condition then shows that the neutrino state $\nu(t)$ is altered as a
result of the variation in density: $\nu(t)$ follows $\nu^m_1$ and
$\nu^m_2$, while the admixtures $\nu^m_1$ and $\nu^m_2$ in $\nu(t)$
undergo only a small change. 

Let us suppose that the neutrinos $\nu_e$ are generated in a region 
with $\rho_0\gg\rho_R$ and then cross layers with continuously
decreasing density $(\rho \rightarrow 0)$. Initially, $\theta^0_m \sim
90^{\circ}$, $\sin^2 2 \theta^0_m \sim \sin^2 2\theta (\rho_R/\rho_0)^2 \ll 1$, 
and $\nu(0)=\nu_e\sim \nu^m_2$. As the density tends to zero, the
state $\nu^m_2$ rotates through $90^{\circ}$, as noted above, and because 
of the adiabatic property, $\nu^m_2$ rotates together with $\nu(t)$.
When $\theta$ is small, $\nu(t)\sim \nu^m_2 \sim \nu_{\mu}$ in the final
state. Thus, $\nu_e$ becomes transformed into $\nu_{\mu}$. As the
distance of the point of creation of the neutrinos from the resonance
layer increases (in $\rho$), $\sin^2 2\theta^0_m$ decreases, and the
difference between $\nu(t)$ and $\nu^m_2$ also decreases. There is an 
attendant reduction in the oscillation depth. In the limit of large
$\rho_0$, the function $\nu(t)$ becomes practically identical with
$\nu^m_2$, i.e., the eigenstate in the medium, and, consequently,
there are no oscillations. This limiting case corresponds to a
nonoscillatory transformation. 

\section{Applications of oscillation effects}

We shall now formulate the general conditions for the above effects
to produce observable consequences.

(1) Both amplification of oscillations and significant changes in the
properties of the $\nu$-beam in the adiabatic regime are due to the
crossing of the resonance layer by the neutrinos. The resonance condition
is satisfied for a particular sign of $l_{\nu}/l_0$ or $\Delta f(0)\cos 2
\theta/\Delta m^2$ [see(8)]. Since replacing $\nu$ with $\bar{\nu}$
produces a change in the sign of $\Delta f(0)$, the resonance condition
is satisfied in a given medium only for a neutrino or an antineutrino.
If the oscillation effects are amplified in the $\nu$-channel, they are
suppressed in the $\bar{\nu}$-channel, and vice versa.

(2) the matter-effect occurs in the charge-asymmetric medium. If the
particle and antiparticle densities are equal, we have 
$\Sigma_i n_i \Delta f_i(0) = 0$.

(3) The typical scale over which the influence of matter is significant
is $l \gtrsim l_0=A/G_{F}\rho$, where $A=3.5 \times 10^4$ km. Hence, it follows
that the thickness of the medium must be greater than 
$d=l\rho >  3.5 \times 10^9 {\rm g/cm}^2$.

These conditions and the adiabatic condition are satisfied in the sun
as well as in the envelopes and cores of collapsing stars.

\section{Neutrino fluxes from collapsing stars}

The effects examined above can occur in the wide range of values of
$\Delta m^2$ and $\sin^2 2\theta$ in the outer layers of the cores
(above the neutrino sphere) and in the envelopes of collapsing stars.

For the purposes of estimates, we shall use the density distribution
in the envelope at the beginning of the collapse, by analogy with the
situation prevailing in white dwarfs:
\be
\rho \approx \rho_0[(R_B/r)-1]^3,
\label{20}
\ee
where $\rho   \simeq (8-10) \times 10^5$ g/cm$^3$ 
and $R_B = 5 \times 10^8$ cm. The 
maximum density in the interior is $\rho_{max}  \simeq 10^9$ g/cm$^3$. For
massive stars, there is, in addition to (20), an extended hydrogen
envelope with $\rho  \approx 10^{-9}$ g/cm$^3$. The envelope may begin to expand 
during the collapse process with velocity $v  \simeq 5000$ km/s.

For collapsing cores, we shall use the model density distribution and
the neutron and electron densities given in Ref. 7. The density in the
neutrino sphere will be taken to be $\rho  \simeq 10^{12}-10^{13}$ g/cm$^3$.

We have used these density distribution to calculate the $\nu$-parameter
ranges (see Fig.2) for which the adiabatic transformation conditions
(1) - (3) of Section 3 are satisfied. The upper limits for $\Delta m^2$
are determined by the maximum density. The lower limits for $\sin^22\theta$
follow from the adiabatic conditions. The values of $\Delta m^2_{max}$
for $\nu_e \leftrightarrow \nu_{\mu}$ and $\nu_e \leftrightarrow \nu_s$ 
($\nu_s$ is the sterile state) are different:
$$
\Delta m^2_{max}(\nu_e \leftrightarrow \nu_{\mu})= 3 \cdot 10^{-2}\Delta
m^2_{max}(\nu_e \leftrightarrow \nu_s).
$$
This difference is largely due to the strong neutronization of the
core $(n_n \gg n_e)$. We note that in the case of the $(\nu_e\leftrightarrow
\nu_s)$-oscillations, the ratio $l_{\nu}/l_0$ changes sign for
$\rho   \simeq 10^9$ g/cm$^3$ in the core. Here, we have a cancellation of the effects
on neutrons, on the one hand, and on electrons and protons, on the other.
This means that, if most of the conversion in the core is for the neutrinos,
most of the conversion in the envelop will be for the antineutrinos.

Since, under the above conditions, the neutrino state is practically
the same as one of the eigenstates in the medium, there is no loss of
coherence due to the spreading of the wave packets corresponding to the 
different $\nu^m_i$.

\begin{figure}[htb]
\centering
\hspace*{-0.60cm}
\includegraphics[width=95mm,angle=00]{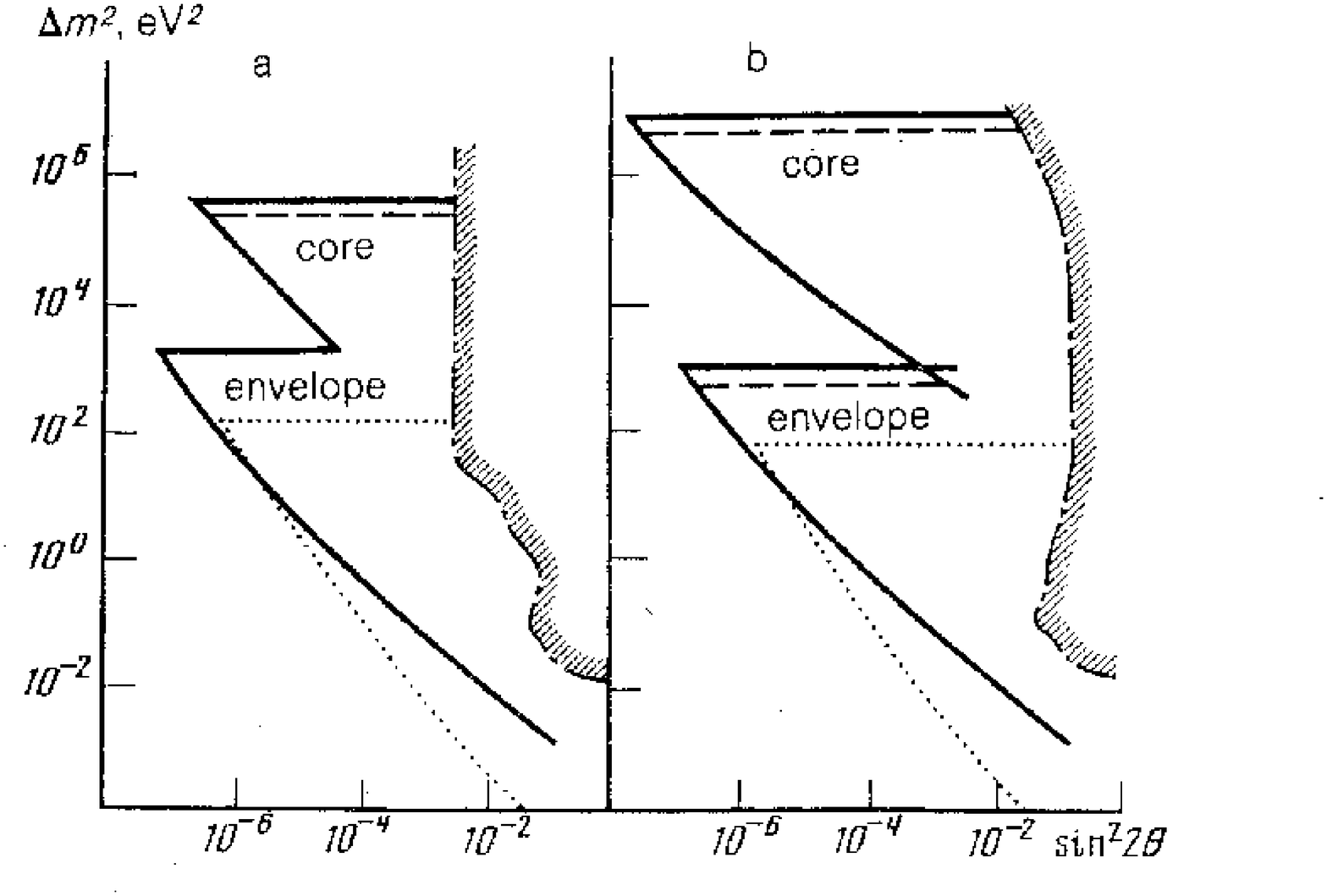} 
\caption{Range of $\nu$-parameters in which the cores and envelopes of collapsing
stars exhibit strong adiabatic transformation: a).  $\nu_e\leftrightarrow
\nu_{\mu}$, b).  $\nu_e \leftrightarrow \nu_s$ oscillations. Below the shaded
line-effect at exit from the envelope: 
$\bar{P}_0\langle(\sin^2\theta)/2$, dotted line - effect 
of an expanding envelope at the end of  the $\nu$-burst
$t=10s)$; dot-dash lines - experimental limits.}
\label{figg2}
\end{figure}

For the ranges of values of the parameters $\Delta m^2, \sin^2 2\theta$ given
in Section 3 (Fig. 2), the factor representing the suppression of the flux
of neutrinos of the original type approaches something between 
$(1/4)\sin^2 2\theta$ and $(1/2)\sin^2 2\theta$.

We must now consider the consequences of the oscillation effect in matter
from the point of view of detection of $\nu$-bursts due to collapses.

(a) Suppose that conditions (1)-(3) are satisfied in the 
$\bar{\nu}_e \leftrightarrow \bar{\nu}_a$ channel, 
where $\bar{\nu}_a = \bar{\nu}_{\mu}$ or $\bar{\nu}_{\tau}$. 
In that case, the $\bar{\nu}_e$ flux 
will be almost completely transformed into the $\bar{\nu}_a$ flux, and vice versa; $\bar
{\nu}_e$ and $\bar{\nu}_a$ will exchange their spectra. It is assumed that
the $\bar{\nu}_{\mu}$ and $\bar{\nu}_{\tau}$ have harder spectra than the
$\bar{\nu}_e$ (the total fluxes are roughly equal for the entire burst).
The exchange of the spectra between the $\bar{\nu}_e$ and $\bar{\nu}_a$
means that there is a substantial increase in the number of events in
scintillation counters detecting $\bar{\nu}_e (\sigma_{\nu}\sim E^2_{\nu})$
(Ref. [8]).

(b) If the resonance condition is satisfied for the neutrinos, $\nu_e
\rightarrow \nu_a$ $(\nu_a = \nu_{\mu},\nu_{\tau})$, an effect analogous to that
just described may be expected in systems based on radiochemical methods
($\nu_e$ detection).

The $\nu_a \leftrightarrow \nu_e$ or 
$\bar{\nu}_a\leftrightarrow \bar{\nu}_e$ 
spectrum exchange in the stellar core leads to a substantial release of
energy in the envelope because of $\nu_e e-$ or $\bar{\nu}_e e-$ scattering.
This may play a definite part in the mechanism responsible for the shedding
of the envelope.

(c) Resonance in the $\nu_e\rightarrow\nu_s$ or $\bar{\nu}_e\rightarrow \bar
{\nu}_s$ channels. The sterile neutrino fluxes appear to be very low. The
effect due to the precession of spin in the magnetic field is negligible for
$H\lesssim 10^{12}$ G and $\Delta m^2 \lesssim 10^3 {\rm eV}^2$. The 
$\nu_e \rightarrow \nu_s$ or
$\bar{\nu}_e\rightarrow{\nu}_s$ transformations will therefore give rise to
strong (by several orders of magnitude) suppression of the $\nu_e$ or $\bar
{\nu}_e$ fluxes.

We emphasize the importance of simultaneous experiments on the detection of
$\nu_e$ and $\bar{\nu}_e$.

We note that, since $l_{\nu}/l_0$ has opposite signs in cores and envelopes, 
a strong effect may be present for both $\bar{\nu}_e$ and $\nu_e$ with
different energies.

In the case of mixing of three or more neutrino types, it is possible that
two or more differences $\Delta m^2$ will fall into the range of strong
adiabatic conversion. A combination of the above effects may then be observed.

We note that matter effects may not remain constant in the course of a $\nu$-
burst $\Delta t\approx 5-20$ s). This time dependence will reflect the
variation in the structure of the core and envelope.\\

\newpage

The authors are indebted to L. Wolfenstein, G. T. Zatsepin, A. Yu. Ignat'ev, 
D. K. Nadezhin, V. A. Rubakov, V. G. Ryasni, and M. E. Shaposhnikov for useful
discussions.


\vspace{0.5cm}

Translated by S. Chomet\\

\vspace{11cm}


\section*{Comments (June 2007)}

1. This paper presents, in particular, our first analytic results on  
the adiabatic conversion of neutrinos in matter.   
It has been written in summer-fall 1985.
In attempt to avoid  problems with publication (we had before), 
we tried to hide the term ``resonance'', and
did not discussed applications to the solar neutrinos; 
also  we have not included references to our previous papers
on the resonance enhancement of neutrino oscillations.

This short paper has been submitted to JETP Letters in the fall 1985 and 
successfully ... rejected.
It was resubmitted to JETP in December of 1985.
The results of the paper have been reported at the 6th Moriond workshop in January
1986 and included in several later reviews. 
The paper was reprinted in ``Solar Neutrinos: The first Thirty Years'',
Ed. J. N. Bahcall, et al., Addison-Wesley 1995.\\

2. The differential equation of third order for the survival probability 
$P$, Eq. (1), has been derived from 
a system  of three differential equations for $P$, 
$R \equiv  Re \langle \nu_e | \nu_\mu \rangle$, and 
$I \equiv  Im \langle \nu_e | \nu_\mu \rangle$. 
The system of equations has been obtained in our first paper: 
Sov. J. Nucl. Phys. {\bf 42},  913 (1985), 
(Yad. Fiz. {\bf 42},  1441 (1985)).\\


3. Analytic results of sec. 3,
have been derived neglecting 
the high order derivatives $d^3P/dt^3$ and $d^2P/dt^2$
in Eq. (1) which is implied by the adiabatic condition.
The resulting equation, 
$$
M (M^2 + 4\bar{M}^2) \frac{d P}{dt} - 
 2\bar{M}^2 \frac{dM}{dt} (2P-1) = 0,
$$
can be easily integrated:
$$
P = \frac{1}{2} + \left(P_0 - \frac{1}{2}\right) \frac{\sqrt{n^2_0 + 1}}{n_0} 
\frac{n}{\sqrt{n^2 + 1}}. 
$$
With the  initial condition
$$
P(0) = 1 - \frac{1}{2}\sin^2 2 \theta_m^0 = 1 - \frac{1}{2(n_0^2 + 1)}
$$
it leads to the adiabatic conversion formula (16).
Noticing that
$$
\frac{n}{\sqrt{n^2 + 1}} = \cos 2\theta_m, ~~~
\frac{n_0}{\sqrt{n^2_0 + 1}} = \cos 2\theta_m^0
$$
one realizes immediately that Eqs. (16) and (18) coincide with
the adiabatic formulas that usually appear in literature.

\end{document}